\documentclass[aps, phsrev, 10pt, twocolumn,superscriptaddress, floatfix]{revtex4-2}

\usepackage{graphicx,color}
\graphicspath{{./Figures/}} 
\usepackage{amssymb}
\usepackage{amsmath}
\usepackage{amsfonts,amsthm} 
\usepackage{natbib}
\usepackage[colorlinks=true,linkcolor=blue,citecolor=blue,urlcolor=blue]{hyperref}
\usepackage[normalem]{ulem}
\usepackage{physics}

\begin{document}

\title{Artificial spin ice on the CaVO lattice: a tricritical bridge from square to kagome}

\author{F. Museur}
\email[Contact: ]{flavien.museur@psi.ch}
\affiliation{Laboratory for Mesoscopic Systems, Department of Materials, ETH Zurich, Switzerland}
\affiliation{PSI Center for Neutron and Muon Sciences, 5232 Villigen PSI, Switzerland}

\author{G. M. Macauley}
\affiliation{Department of Physics, Princeton University, Princeton, NJ 08540 USA}

\author{S. Nilsson}
\affiliation{Laboratory for Mesoscopic Systems, Department of Materials, ETH Zurich, Switzerland}
\affiliation{PSI Center for Neutron and Muon Sciences, 5232 Villigen PSI, Switzerland}

\author{V. Scagnoli}
\affiliation{PSI Center for Neutron and Muon Sciences, 5232 Villigen PSI, Switzerland}

\author{P. M. Derlet}
\affiliation{PSI Center for Scientific Computing, Theory and Data, 5232 Villigen PSI, Switzerland}

\author{L. J. Heyderman}
\affiliation{Laboratory for Mesoscopic Systems, Department of Materials, ETH Zurich, Switzerland}
\affiliation{PSI Center for Neutron and Muon Sciences, 5232 Villigen PSI, Switzerland}

\begin{abstract}
Artificial spin ices (ASIs) are arrays of interacting nanomagnets in which phase transitions can be imaged and controlled. Studies of the thermodynamic have focused mainly on the square and kagome lattices, and on their variants. Here we study artificial spin ice on the CaVO lattice, formed of squares and octagons. At fixed lattice constant, the ratio $r$ of the square and octagon link lengths tunes the interactions strongly while changing the geometry only slightly. Magnetic force microscopy and Monte Carlo simulations show a continuous bridge from square-ice-like to kagome-like behavior. At small $r$ the system has a single ordering transition, whereas at large $r$ it passes from a correlated paramagnet to charge order and then long-range order. The two regimes meet in a multicritical region, accurately modeled by a Blume-Capel model of the octagon toroidal moments. From this picture, we design a comparative magnetometry protocol that unambiguously separates interaction effects from single-nanomagnet effects in thermally active samples. Finally, the charge-ordered region maps onto a tunable dimer model, with a Coulomb phase and pinch points along the line between two ordered dimer phases. The CaVO ASI thus provides access to a rich phase diagram only through geometric tuning, without requiring extreme experimental conditions.
\end{abstract}

\maketitle

\begin{figure}[t]
    \centering
    \includegraphics[width=0.9\linewidth]{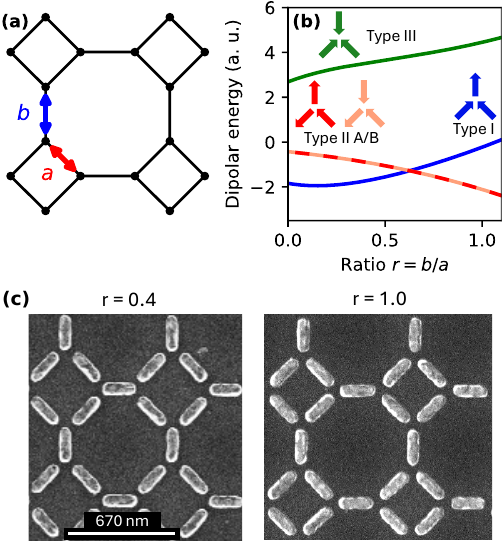} 
    \caption{\textbf{One small geometry change tunes the vertex energies of the CaVO lattice.} (a) Definition of the lattice and of the $a$ and $b$ tuning parameters. (b) Point-dipole energies of the vertex types versus $r=b/a$; Type I and Type II cross at $r=0.63$. (c) SEM images of CaVO arrays with $r=0.4$ (left) and $r=1.0$ (right)}
    \label{figure:1}
\end{figure}

\textcolor{red}{Note: This manuscript is a report on the work presented by F. Museur at several international conferences. This is a preliminary draft of a manuscript and has not yet been reviewed and agreed on by all authors.} 

Artificial spin ices (ASIs) are arrangements of dipolar-coupled nanomagnets, with magnetic moments that behave like Ising spins as a result of shape anisotropy. They have been successfully implemented as platforms for imaging and controlling phase transitions in systems with long-range interactions~\cite{skjaervoAdvancesArtificialSpin2019}. 
There is an intimate link between the geometry of an artificial spin ice, which determines the intermagnet interactions, and the collective phenomena exhibited. In particular, the nanomagnet size and shape~\cite{zhangArtificialMagneticTripod2023, chaurasiyaControllingDegeneracyMagnetization2022}, the lattice geometry~\cite{talapatraMagneticTunabilityPermalloy2020} and the lattice parameter~\cite{hofhuisGeometricalFrustrationCompeting2021} all determine the behaviour of an artificial spin ice.

Studies of thermally driven phase transitions in artificial spin ice have focused mostly on two lattices: square ice, with nanomagnets on the edges of a square lattice, and kagome ice, with nanomagnets on the edges of a hexagonal lattice. The two behave in very distinct ways. Artificial square ice orders readily into an antiferromagnetic, two-in/two-out ground state, through a transition in the two-dimensional Ising universality class. The ground state is easily detectable via MFM measurements~\cite{wangArtificialSpinIce2006a}, and thermally active samples have been shown to possess the correct critical properties close to the transition~\cite{sendetskyiContinuousMagneticPhase2019}. However, in order to reach a true spin liquid state, the degeneracy between the two lowest energy vertex types needs to be restored, for example by elevating one sublattice with respect to the other~\cite{perrinExtensiveDegeneracyCoulomb2016}, at the cost of a more complex fabrication process.

Artificial kagome ice, in contrast, is highly frustrated, and its predicted phase diagram is richer~\cite{mollerMagneticMultipoleAnalysis2009,chernTwoStageOrderingSpins2011}. On cooling, it first crosses over from a paramagnet to a spin liquid without long-range order. It then orders into a phase where the magnetic charges on the vertices crystallize antiferromagnetically while the spins stay disordered. This peculiar phase, resulting from the second nearest-neighbor interactions, was used as a building block of the concept of magnetic moment fragmentation~\cite{brooks-bartlettMagneticMomentFragmentationMonopole2014,canalsFragmentationMagnetismArtificial2016,lhotelFragmentationFrustratedMagnets2020,museurFerromagneticFragmentedState2026}. In this field-theoretical framework, each magnetic moment can be decomposed into the sum of a divergence-full fragment and a divergence-free fragment by applying the Helmholtz decomposition. The dipolar interactions in artificial kagome spin ice ensure that the divergence-full fragment remains long-range ordered, while the divergence-free fragment fluctuates over an almost flat energy landscape, carrying a residual entropy. Furthermore, this fragment exhibits dipolar correlations, signaling that it is a Coulomb phase; therefore, it can be elegantly mapped to a dimer model. The fragmentation formalism highlights a fundamental connection between many spin systems with long-range interactions, and a dimer system with hard-core repulsive interactions. In artificial kagome ice, however, this charge order has proved hard to reach. It is driven by second-neighbor interactions, so sets in well below the blocking temperature of the nanomagnets, and the moments freeze before the charges can order~\cite{rougemailleArtificialKagomeArrays2011,canalsFragmentationMagnetismArtificial2016}.

Together with the triangular lattice, the above lattices are the Euclidean lattices, formed of only one shape paving the plane. From an artificial spin ice standpoint, they are simple systems in that all vertices are identical. However, there exist many ways to pave the plane with more than one shape while still keeping all vertices the same: the resulting lattices are called Archimedean. A detailed numerical investigation of the physics of out-of-plane ASIs on such lattices was first conducted in~\cite{pacMagneticOrderingOutofplane2026}. It motivated the investigation of a more standard in-plane ASI on the Ruby lattice~\cite{berchiallaRealizingBlumeCapelDegrees2026}. The authors realize that this ASI could support a continuous tuning of its geometry, by compressing some of the shapes of the lattice and expanding others. Although it did not change the ground state of the lattice, it resulted in a significant change in the energy levels at the vertex level and in the thermodynamic pathways to reach the ground state.

\begin{figure}[t]
    \centering
    \includegraphics[width=0.9\linewidth]{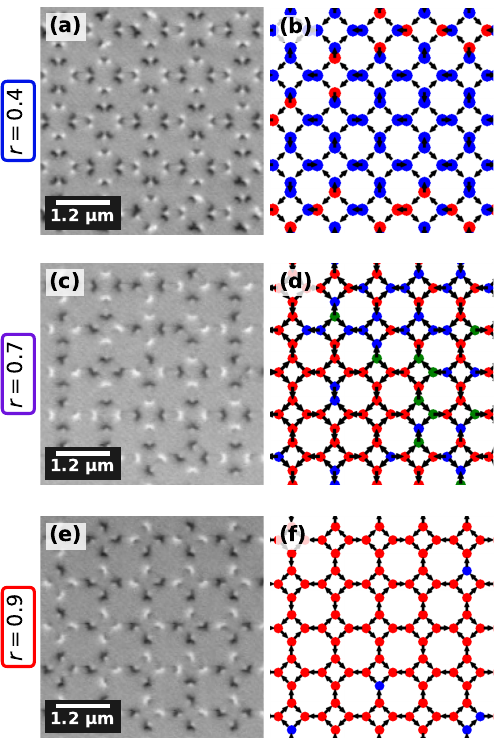}
    \caption{\textbf{The dominant vertex type in as-grown arrays changes with $r$.} MFM contrast (left) and extracted spin maps (right) for $r=0.4$ (a-b), $0.7$(c-d) and $0.9$ (e-f). Vertex colours: blue, Type I; red, Type II; green, Type III. Scale bars, 1.2~$\mu$m.}
    \label{figure:2}
\end{figure}

In this work, we study artificial spin ice on another Archimedean lattice, the CaVO lattice, and show that a single geometric ratio tunes it continuously from square-like to kagome-like behavior. Called the CaVO lattice, it is formed of squares and octagons (see Fig.~\ref{figure:1}a), and derives its name from the CaV$_4$O$_9$ compound of similar structure~\cite{starykhOriginSpinGap1996}. Each vertex has a low connectivity of three and is surrounded by one square and two octagons. We place Ising nanomagnets on the links of such lattices. Similarly to the kagome lattice and square lattices, and in contrast to the ruby lattice, the CaVO lattice is bipartite, meaning that it can support the charge ordered phase described above. Monte Carlo simulations show that the two regimes combine to exhibit a tricritical behavior, well described by a Blume-Capel model, with a Coulomb phase along the transition line between them. Magnetic force microscopy and SQUID magnetometry on arrays across this range confirm small changes in ratio results in large changes in the dominant vertex type as well as the low-temperature dynamics.

To reach the charge-ordered phase, we tune the vertex energies by changing the relative sizes of the square and octagonal plaquettes, as was done for the ruby lattice~\cite{berchiallaRealizingBlumeCapelDegrees2026}. As the squares shrink, the octagon links between them grow, so the lattice constant $b+\sqrt{2}a$, the distance between the centers of neighboring squares, stays fixed (Fig.~\ref{figure:1}c). This transformation preserves the lattice symmetries, and minimizes changes in further neighbor couplings. Each lattice is then defined by the ratio $r=b/a$.

Changing $r$ reorders the vertex energies (Fig.~\ref{figure:1}b). At low $r$, Type I vertices have the lowest energy and the ground state is unique. At high $r$, the two degenerate Type II subtypes, A and B, are lowest. At $r=0.63$, Type I, IIA and IIB have equal energy. A minute change in geometry thus moves the lattice from a unique ground state, through a triple degeneracy, to a double degeneracy, all within a single geometry, which is rare among artificial spin ices.

\begin{figure}[t]
    \centering
    \includegraphics[width=\linewidth]{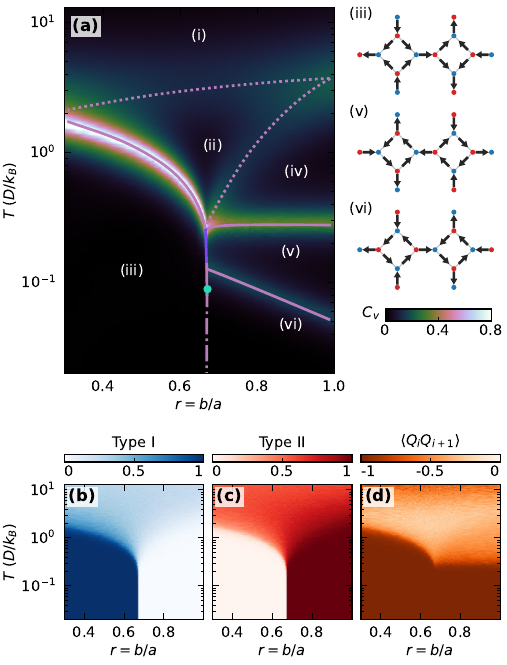}
    \caption{\textbf{Phase diagram from Monte Carlo simulations.} (a) Specific heat $C_v$ as a function of the ratio $r=b/a$ and temperature $T$, with regions (i)--(vi) discussed in the text. Solid lines: continuous transitions, with the (iii)--(v) line in purple; dot-dashed line: discontinuous transition; dotted lines: crossovers; green dot: tricritical point. Right of (a), from top to bottom: spin configurations of the three low-temperature phases over two unit cells, taken from Monte Carlo snapshots at the stated $r$ and $T$ and drawn on the $r=0.7$ lattice (same lattice constant) with arrows of equal length. Blue and red dots mark positive and negative vertex charges. In (iii) (snapshot at $r=0.41$, $T=0.05\,D/k_\mathrm{B}$) all vertices are Type I. In (v) ($r=0.86$, $T=0.15\,D/k_\mathrm{B}$) all vertices are Type II and the squares close their flux in random directions. In (vi) ($r=0.86$, $T=0.03\,D/k_\mathrm{B}$) all squares close their flux in the same direction. Below the snapshots: color scale of (a); the color scales of (b)--(d) are above each map. (b) Population of Type I vertices. (c) Population of Type II vertices. (d) Nearest-neighbor charge correlator $\langle Q_iQ_{i+1}\rangle$. Populations and correlator use bulk vertices only; each map averages three neighboring temperatures.}
    \label{figure:3}
\end{figure}

Magnetic force microscopy (MFM) of as-grown arrays confirms that $r$ selects the dominant vertex type (Fig.~\ref{figure:2}). The arrays have a lattice constant of 1.2~$\mu$m and $400\times100\times20$~nm$^3$ permalloy nanomagnets (see Sec.~\ref{sm:fabrication} of the Supplemental Material for methods). For low ratio ($r=0.4$, left-most column), we find a very large population of Type I states, leading to a ground state  where all octagons are flux closed in a staggered fashion. This ground state extends over large areas of the as-grown arrays (Fig.~\ref{figS:mfm}). This state is very reminiscent of the ground state of square ice, where all squares are flux closed in a staggered way as well. At high ratio ($r=0.9$, right column), almost all vertices are Type II. Every square is then flux-closed. Since Type IIA and IIB have equal energy, however, the macrospins on the squares close their flux in random directions, and the octagon-link macrospins between them stay disordered. For a ratio close to the degeneracy point ($r=0.7$, center column), we observe a more even mixture of all type of vertices. In this degenerate state, we can expect further correlations beyond nearest-neighbor to become more prominent.


Such a collection of different vertex degeneracies in the energy diagram warrants a more detailed characterization of the resulting thermodynamic phases. Monte Carlo simulations with the full dipolar interaction show two ordering routes, at low and high $r$, separated by a multicritical region (see the specific heat map in Fig.~\ref{figure:3}a). At high temperature, the system is paramagnetic and all vertex types are present with non zero probability (region (i)). Around a temperature scale of $T \approx 3$, the high energy Type III vertices are suppressed. For $r<0.67$, the system behaves like square ice. As the temperature is lowered, the system experiences a single transition to a long range ordered state (region (iii)) formed of only Type I vertices (see Fig.~\ref{figure:3}b). As the symmetry breaking concerns the direction of flux-closure of a particular octagon, this transition is of the usual Ising 2D class. As observed in the MFM contrast images (see Fig.~\ref{figure:2}a and b), every octagon is flux-closed in a staggered way. For $r>0.67$, the system instead follows the kagome sequence. On cooling, the system first experiences a crossover to region (iv) with only Type II A/B vertices (see Fig.~\ref{figure:3}c), meaning that each square is flux-closed but in a random direction. It is indicated by a broad feature in the specific heat at high temperatures. It is then followed by a phase transition to a charge ordered phase, visible as the horizontal line between regions (iv) and (v) at $T \approx 0.3$. This phase is charge-ordered, similar to the KII phase of kagome spin-ice. The vertices stay of Type II type, but the magnetic charges that they carry order antiferromagnetically such that "two-in / one-out" vertices are only neighbor to "one-in / two-out" vertices and reciprocally. In a spin language, the sublattice of octagonal spins, which is a square lattice of the same constant as the CaVO lattice, orders following the usual "two-in / two-out" square ASI order. This explains why the temperature of this transition is independent on the ratio: it is mediated by the second nearest-neighbor interactions, which are kept mostly constant by our geometrical transformation. In Fig.~\ref{figure:3}d, we plot the charge crystallization parameter, which shows that region (iii) also exhibits full charge order. At still lower temperature, the squares choose a common direction of flux closure (region (vi)), through third and further neighbor interaction. The CaVO lattice therefore links square-like and kagome-like behavior through a single parameter. 

\begin{figure}[t]
    \centering
    \includegraphics[width=\linewidth]{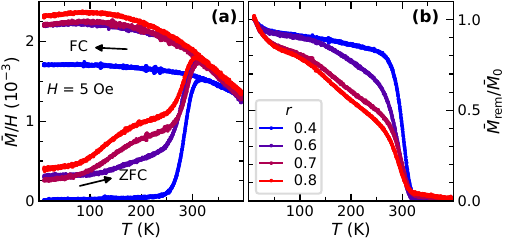}
    \caption{\textbf{Arrays with higher $r$ fluctuate more below the blocking temperature.} (a) ZFC-FC susceptibility in 5~Oe and (b) remanent magnetization  normalized to its low-temperature value following a field applied in the (10) direction. The same $r=0.4$, $0.6$, $0.7$ and $0.8$ sample series is shown in both plots. Curves are corrected for remanent field and a sample-dependent Py impurity (see Sec.~\ref{sm:impurity} and Figs.~\ref{figS:protocol}--\ref{figS:correction}).}
    \label{figure:4}
\end{figure}

The competition between flux closure of squares and octagons results in a multicritical region in the center of the phase diagram. Close to the critical ratio of $r\approx 0.63$, we model this competition with an effective Hamiltonian of the toroidal moments on each octagon plaquette. We define an effective variable $S_i = 0, \pm1$ at the center of each octagon, which is $+1$ if the spins flux-close anticlockwise, $-1$ clockwise, and $0$ in the other cases. At a vertex level, for a ratio greater much greater than $0.63$ (see Fig.~\ref{figure:1}), Type II vertices are preferred and no octagons are flux-closed at a nearest-neighbor level. This is confirmed in the $r = 0.9$ MFM image in Fig.~\ref{figure:2}e and f. However for r less than $0.63$, vertices of Type I are preferred and all octagons are necessarily flux closed, as is evident in Fig.~\ref{figure:2}a and b. The ratio r therefore acts like an effective anisotropy term ($\Delta \sum_i S_i^2$ with $\Delta \propto r-0.63$ at leading order). This picture only includes the nearest-neighbor interactions; but the longer-range part of the dipolar interactions also favors flux closure. In particular, if we place ourselves at the degeneracy point where Type I and Type II energy are equal, the energetics of the system are controlled by the second and farther nearest neighbors interactions. A point dipole calculation then shows that the lowest energy state is the same as the low r case, with every vertex of Type I and all octagons flux closed.  It corresponds to an antiferromagnetic $S_i = \pm1$ order of the toroidal moment. Therefore, the long range part of the Hamiltonian acts like an small effective antiferromagnetic Ising term $J \sum_{i,j \text{ nn}} S_iS_j$ with $J>0$. Putting both terms together, the competition of vertex types can thus be modeled as a Blume-Capel Hamiltonian:
\begin{equation}
    \mathcal{H} = J \sum_{i,j\text{ nn}} S_iS_j + \Delta \sum_i S_i^2
\end{equation}
which is the simplest model to exhibit a tricritical point (shown as a green dot in Fig.~\ref{figure:3}a, and in the zoom of Fig.~\ref{figS:hist}). There, the nature of the transition between the ordered $S=\pm 1$ and $S=\pm 0$ turn form continuous (full line) to discontinuous (dot-dashed line). Further more, the mapping explains an otherwise surprising difference in the critical ratios between the dipolar energy calculation in Fig.~\ref{figure:1} ($r \approx 0.63$) and Monte-Carlo simulations ($r \approx 0.67$). The competition between exchange and anisotropy inherent to this model make it so that the critical point is located at $\Delta / J \approx 0.5$ instead of $0$. Previous artificial spin ices that were modelled as a system of toroidal moments could only see one side of the Blume-Capel phase diagram~\cite{berchiallaRealizingBlumeCapelDegrees2026,yueToroidicPhaseTransitions2024}; the CaVO ASI is the first known experimental example where the full tricritical character of the Blume-Capel model can be realized. Such physics appears to be a more generic feature of dipolar frustrated magnets: in pyrochlore spin ices~\cite{rabanMultipleSymmetrySustaining2019,brooks-bartlettMagneticMomentFragmentationMonopole2014}, the competition between ferromagnetic interacting moments and an antiferromagnetic internal field was shown to be modeled by a $S=2$ Blume-Capel Hamiltonian. However, observing the first order transitions experimentally require extreme conditions: very low temperatures, high pressures and/or high magnetic fields. The CaVO artificial spin ice, thanks to its tunable geometry, is therefore an ideal system to explore tricritical physics.

\begin{figure}[htb]
    \centering
    \includegraphics[width=0.95\linewidth]{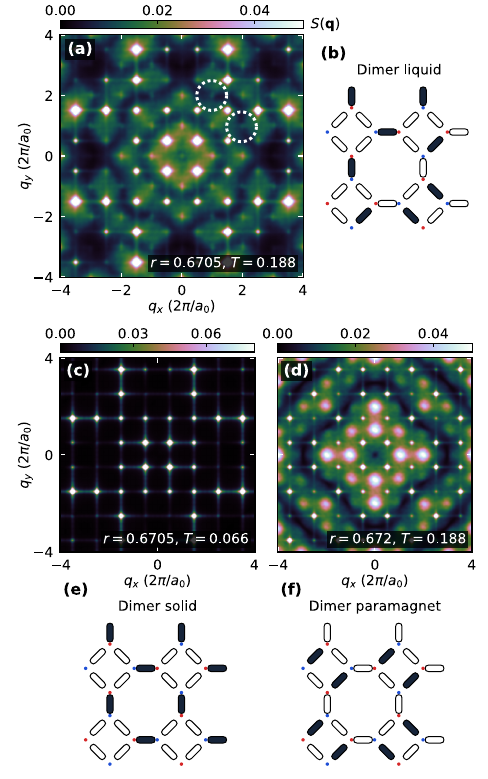}
    \caption{\textbf{A Coulomb phase on the transition line between the dimer solid and the dimer paramagnet.} (a) Magnetic structure factor $S(\mathbf{q})$ at $r=0.6705$, $T=0.188\,D/k_B$, on the transition line between regions (iii) and (v): pinch points (in the center of the dashed circles), characteristic of algebraic dipolar correlations, coexist with bright charge-order Bragg peaks. (b) Dimer liquid: dimers occupy octagon links and square edges with equal probability. (c) Dimer solid, deep in region (iii) ($r=0.6705$, $T=0.066\,D/k_B$): Bragg peaks only. (d) Dimer paramagnet, region (v) ($r=0.672$, $T=0.188\,D/k_B$): charge-order Bragg peaks on a diffuse background. (e) In the dimer solid, every octagon link holds a dimer. (f) In the dimer paramagnet, each square holds its two dimers in one of two equivalent orientations, chosen at random. $S(\mathbf{q})=|\mathbf{M}_\perp(\mathbf{q})|/N$ is computed from the 1014 bulk spins of a $17\times17$ lattice and smoothed with a Gaussian of width $0.039\times 2\pi/a_0$; color scales are reduced in (a) and (d) to show the diffuse scattering.}
    \label{figure:5}
\end{figure}

Because only the interactions change with $r$, while the individual nanomagnets, the lattice constant and the lattice symmetry stay fixed, a series of CaVO arrays can isolate the effect of interactions from single nanomagnet blocking in bulk magnetic measurements. We tested this with standard magnetometry, (performed on a commercial MPMS3 SQUID-VSM from Quantum Design) on thermally active arrays with $r=0.4$, $0.6$, $0.7$ and $0.8$. For these measurements, we manufactured a series of CaVO ASIs covering an area of $3\times 3$ mm$^2$ with identical nanomagnets but different ratios. In order to ensure that the nanomagnets are thermally active around room temperature, they are smaller than the ones typically designed for MFM measurements with dimensions of about $185 \times 65 \times 3.2$ nm (see Fig.~\ref{figure:1}c). In particular, the thickness, the quality of the evaporated film and the quality of the patterning all have a very large impact on the thermally active behavior (Sec.~\ref{sm:thickness} and Figs.~\ref{figS:thickness}--\ref{figS:py2603}).

We ran ZFC-FC protocols with a low magnetic field and remanent magnetization measurements to characterize both the energy barriers associated with thermal fluctuations and the ordering. Since the nanomagnets are identical for all samples, we can expect the similarities in the data to be explained by single nanomagnet properties, while the differences must be caused by variations in the intermagnet interactions. The main common features in all samples are the meeting of the ZFC-FC curves and the collapse of the remanent magnetization, in Fig.~\ref{figure:4}a and b respectively, which both occur at about 310~K. We associate these features with the onset of superparamagnetism in all samples, with this temperature corresponding to the blocking temperature. However, there are significant differences between the samples below the blocking temperature. In particular, in Fig.~\ref{figure:4}a, the ZFC r = 0.4 curve (in blue) remains at 0 magnetization while, when increasing r, a bump in the curves appears. It becomes bigger as r is increased, and is associated with an increasing dip in the remanent magnetization curves in Fig.~\ref{figure:4}b. The FC magnetization is also significantly lower for r = 0.4 sample than any of the other ones. 

This effect can be explained when taking into account the different correlations expected for different ratios. CaVO ASIs with high ratios (r = 0.7, 0.8, in burgundy and red respectively) have a large population of vertices of Type II, which is degenerate between A and B variants. Then, under a small external field like in a ZFC protocol, the response of the system can be larger as switching macrospins between the two variants does not cost energy. On the contrary, the sample with low ratio (r = 0.4) will have a large population of vertices of type I. These are not degenerate, so switching macrospins will either create vertices of Type II or III (which cost more internal energy), or require switching an entire patch of octagons (the probability of which is vanishingly low). When relaxing from a state magnetized at saturation (which has a mixture of type I and II on different sites), a similar effect occurs. Getting to the single state with only Type I vertices requires going through many intermediate states with higher energy, whereas getting to one of the many states with only Type II vertices can be done by only lowering the energy at each macrospin flip. This picture also explains why the r = 0.7 curve appears so similar to the r = 0.8 one, but not the r = 0.6 one (in purple). Despite all samples being geometrically close, the thermodynamics of r = 0.7 is much closer to r = 0.8, with a preponderance of vertices Type II, while r = 0.6 is still dominated by vertices of Type I. It is precisely the tricritical nature of the system, with a paramagnetic phase separated from an ordered phase by a transition that becomes first-order, that ensures a large jump in observables from just a small shift in the phase diagram.

From these observations, the conclusions are twofold. First, the differences in interactions have predictable effects on the magnetization as a function of temperature. Second, a single blocking temperature cannot fully explained the measured data, given that there remain significant dynamics below the main blocking temperature. To refine the thermal model of our system, it is necessary to introduce a second blocking temperature, at an energy scale of about 50 K, but with a very small density of states. It corresponds to a rare, but not impossible, pathway for macrospins to fluctuate at temperatures much below the main blocking temperature of about 310~K. This same low energy scale has been identified in a previous work as originating from the oxidation of the edge of the nanomagnets, which can provide pinning sites for the magnetization to slip at a lower energy cost~\cite{morleyVogelFulcherTammannFreezingThermally2017}.Together with the effect of varying interactions, this provides a complete picture of magnetometry measurements on out-of-equilibrium magnetic metamaterials.

The multicritical region near $r=0.7$, however, remains out of reach in these arrays: the charge-ordering transition lies below the main blocking temperature, so the moments freeze before the charges order. 
Any phase exhibiting charge order on such a bipartite lattice is of particular interest, because on a bipartite lattice any charge-ordered state does so through moment fragmentation~\cite{brooks-bartlettMagneticMomentFragmentationMonopole2014,canalsFragmentationMagnetismArtificial2016}. This mapping works both for fully and partially ordered states, meaning that the entire region of the phase diagram showing full charge crystallization (regions (iii), (v) and (vi) in Fig.~\ref{figure:3}a) can be thought of as representing a tuneable dimer model (see Sec.~\ref{sm:fragmentation} and Fig.~\ref{figS:frag}). Dimer models have long been a staple of statistical physics, as they can be solved exactly using standard mathematical methods~\cite{wuDimersTwodimensionalLattices2006,nagleDimerModelsAnisotropic1989} while still showing a very rich behavior. The CaVO ASI is a promising candidate to realize tuneable dimer physics experimentally. Region (iii) of the phase diagram can be interpreted as a dimer solid, where all octagonal sites are occupied by a dimer. Region (v), by contrast, is a dimer paramagnet: the magnetic charges stay ordered, but because the squares close their flux in random directions, each square holds its dimers in one of two equivalent ways (Fig.~\ref{figure:5}f).

Right between these two phases, along the transition line between regions (iii) and (v) (drawn in purple in Fig.~\ref{figure:3}a), is where a true Coulomb phase can survive. Here, the vertex energetics are balanced so that dimers have the same probability to occupy every link of the lattice (see Fig.~\ref{figure:5}b), and magnetic correlations are algebraic by definition. On this transition line, the magnetic structure factor plotted in Fig.~\ref{figure:5}a shows the well known pinch-point patterns, characteristic of dipolar correlations, as well as Bragg peaks indicating charge order. The fact that pinch-points are only observed on this transition line emphasizes the fact that Coulomb phases are critical in nature~\cite{henleyCoulombPhaseFrustrated2010}. The kagome dipolar spin ice, where the pinch point phase exists in a large fraction of the phase diagram, appears as a remarkably special example. 

The CaVO lattice links artificial square and kagome ice in one geometry. A single ratio $r$ moves it from a square-like ordering transition, through a tricritical point described by a Blume-Capel model, to a kagome-like sequence of paramagnetic spin liquid, charge order, then full order. The competition between different orders has very strong consequences on the thermodynamics, that can be clearly detected with standard magnetometry even in partially out-of-equilibrium samples. Along the transition line between the two regimes, the system realizes a Coulomb phase, making the lattice a tunable realization of a dimer model. Further geometrical variants, like rotating the squares or compressing them into rhombus shapes, can break the degeneracy between vertices of Type IIA and IIB, creating even more low-temperature phases. The CaVO ASI brings them all within simple geometric adjustments, something that in bulk spin systems typically require extremes of temperature, magnetic field or pressure. 


\begin{acknowledgments}
    \textemdash
\end{acknowledgments}
%

\bibliography{biblio}

\onecolumngrid
\clearpage

\begin{center}
\textbf{Supplemental Material}
\end{center}


\setcounter{section}{0}
\renewcommand{\thesection}{S-\Roman{section}}

\setcounter{equation}{0}
\renewcommand{\theequation}{S\arabic{equation}}
\renewcommand{\theHequation}{S\arabic{equation}} 

\setcounter{figure}{0}
\renewcommand{\thefigure}{S\arabic{figure}}
\renewcommand{\theHfigure}{S\arabic{figure}}

\setcounter{table}{0}
\renewcommand{\thetable}{S\arabic{table}} 
\renewcommand{\theHtable}{S\arabic{table}}
\setcounter{page}{1}
\renewcommand{\thepage}{S\arabic{page}} 

\makeatletter

\makeatother

\section{Sample fabrication}
\label{sm:fabrication}

\paragraph*{Process.}
The arrays were made by electron-beam lithography, thermal evaporation of permalloy (Py, Ni$_{80}$Fe$_{20}$) and lift-off, in the Laboratory for Mesoscopic Systems and the PSI cleanroom, following the route used for earlier thermally active artificial spin ices~\cite{sendetskyiMagneticDiffuseScattering2016,sendetskyiContinuousMagneticPhase2019}. The steps are as follows.
(1)~A Si substrate is rinsed in water, sonicated for 3~min in acetone and 3~min in isopropanol (IPA), dried with N$_2$ and dehydrated on a hot plate. PMMA resist (950K, 2\% in ethyl lactate) is spin-coated and baked for 2~min at 175~$^\circ$C.
(2)~The patterns are exposed with a Vistec EBPG 5000PlusES electron beam writer with 100 keV accelerating voltage. Each lattice is generated with a custom Python lattice builder, which writes main field of $500\times500~\mu$m$^2$ with 4.5~$\mu$m sub-fields with a quicker fracturing than the base RAITH Beamer software. This enables us to expose a chip with 4 ratios, each having 36 main fields (so a 3x3 mm$^2$ area) in 1h30 instead of 5 hours.
(3)~The resist is developed in MIBK:IPA 1:3.
(4)~Py is deposited by thermal evaporation in a Balzers evaporator and capped with Al to prevent oxidation. The base pressure was $6\times10^{-6}$~mbar and the evaporation rate about 0.5--1~\AA/s.
The Al cap is 2~nm on the films studied by MFM, and 5--6~nm (quartz-monitor reading) on the SQUID samples.
(5)~The resist and the metal on top of it are removed by lift-off in acetone. The samples first spend 12 hours in acetone, then are sonicated for 20~min in clean acetone, 20 more minutes in clean IPA and dried with N$_2$.

\begin{figure}[b]
\centering
\includegraphics[width=6in]{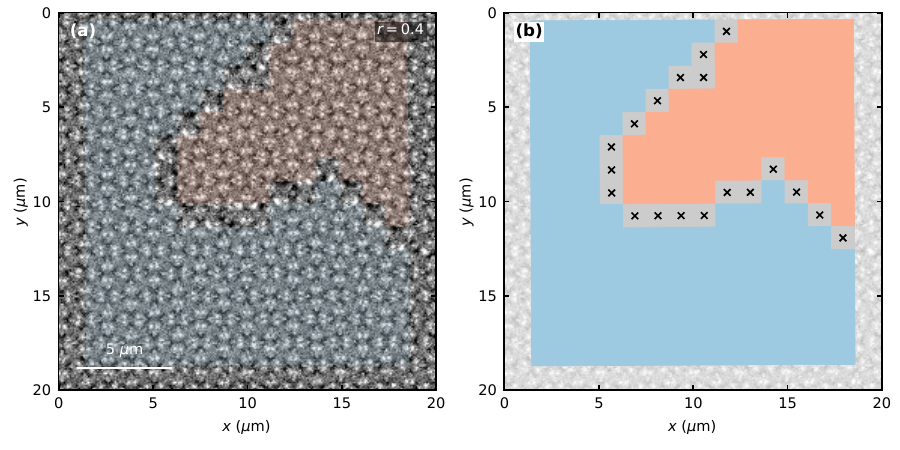}
\caption{\textbf{Extended ground state in an as-grown array.} (a)~Lift-mode MFM phase image of the as-grown $r=0.4$ array (lattice constant $1.2~\mu$m), with the two ground state domains lightly shaded. All vertices inside the domains are of Type I. (b)~Octagon map of the same field. Each tile is one octagon. Light blue and light red tiles are flux-closed octagons of the two degenerate staggered ground states, which differ by reversing every moment. Grey tiles with crosses are octagons that are not flux closed; all of them lie on the wall between the two domains.}
\label{figS:mfm}
\end{figure}

\paragraph*{MFM samples.}
The MFM arrays have a lattice constant $c=1.2~\mu$m for all $r$, nanomagnets of $400\times100$~nm$^2$ and a Py thickness of 20~nm (10~nm in an earlier series). Each array covers $149\times149~\mu$m$^2$, about 124$\times$124 unit cells. The MFM series covers $r=0.3$ to 0.9 in 16 arrays, with finer spacing in geometry around the critical 0.65 - 0.75 ratios. The Py is thick enough that the islands do not flip at room temperature, so the as-grown state is frozen in during growth.

\paragraph*{SQUID samples.}
For magnetometry, an array covers $3\times3$~mm$^2$, which gives about $1.2\times10^8$ nanomagnets for $c=670$~nm.
The islands are about $180\times60$~nm$^2$
and the arrays of different $r$ at a given $c$ share the same islands. A bare Py film ($5\times5$~mm$^2$) is evaporated in the same run. It is used to measure $M_\mathrm{Py}(T)$ (Sec.~\ref{sm:impurity}).

\section{Large-field MFM image}
\label{sm:mfm}

Figure~\ref{figS:mfm} shows that as-grown arrays can hold the staggered flux-closed ground state over large areas. The image is a $20\times20~\mu$m$^2$ lift-mode phase scan (512 pixels per line) of an as-grown $r=0.4$ array with lattice constant $1.2~\mu$m. To assign the moments, we register the ideal lattice on the topography image and fit the high-pass-filtered phase image as a sum of one dipole template per island, a positive and a negative pole at the two island ends, with a free amplitude for each island. All amplitudes are fitted together, so the contrast at a vertex is shared among the three islands that meet there, and the sign of each amplitude gives the moment direction. Of the 960 complete vertices, 96\,\% are Type~I, 4\,\% Type~II and none Type~III. Of the 210 octagons inside the field, 191 are flux closed, and every pair of neighbouring closed octagons circulates in opposite senses, as the staggered state requires. The closed octagons form two domains of the two degenerate staggered states (131 and 60 octagons), and all 19 octagons that are not closed lie on the single wall between them.

\section{Tricritical point}
\label{sm:hist}

\begin{figure}[tb]
\centering
\includegraphics[width=\textwidth]{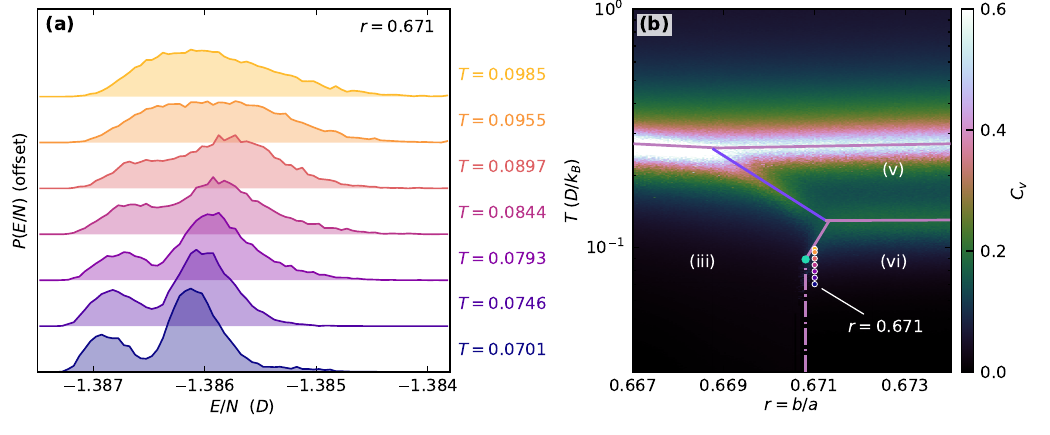}
\caption{\textbf{Tricritical point.} (a) Distribution $P(E/N)$ of the energy per spin at $r=0.671$ for seven temperatures from $T=0.0985$ (top) to $0.0701\,D/k_\mathrm{B}$ (bottom), curves offset vertically; $17\times17$ lattice, 100 replicas, fixed-temperature runs. On cooling, one peak splits into two separated by $\Delta E/N\approx0.8\times10^{-3}\,D$. (b) Specific heat near the critical ratio, $0.667\le r\le0.674$, from temperature sweeps on an $11\times11$ lattice. Lines as in Fig.~\ref{figure:3}a: solid, continuous transitions, with the (iii)--(v) line in purple; dot-dashed, discontinuous transition. Green dot: tricritical point, at $r_t\approx0.6708$, $T_t\approx0.09\,D/k_\mathrm{B}$. Colored dots: the temperatures of (a), at $r=0.671$.}
\label{figS:hist}
\end{figure}

The Blume-Capel mapping of the main text predicts that the transition between the ordered ($S=\pm1$) and disordered ($S=0$) toroidal states changes from continuous to discontinuous at a tricritical point. Figure~\ref{figS:hist}b zooms on the specific heat map around the critical ratio, from temperature sweeps on an $11\times11$ lattice with steps of $4\times10^{-5}$ in $r$. At every temperature between 0.04 and $0.10\,D/k_\mathrm{B}$, the maximum of $C_v$ along $r$ lies at $r=0.6705$--$0.6710$, which fixes the boundary between regions (iii) and (vi). This boundary ends, and the line bends towards the (v)--(vi) line, at the tricritical point $r_t\approx0.6708$, $T_t\approx0.09\,D/k_\mathrm{B}$, which we read from the map to about $\pm0.01\,D/k_\mathrm{B}$.

At a discontinuous transition the energy distribution has two peaks, from the two coexisting phases. Figure~\ref{figS:hist}a shows histograms of the energy per spin $E/N$ at $r=0.671$, from fixed-temperature Monte Carlo runs on a $17\times17$ lattice with 100 independent replicas, at seven temperatures between $T=0.0985$ and $0.0701\,D/k_\mathrm{B}$, marked in Fig.~\ref{figS:hist}b. On cooling below $T_t$, the single peak splits into two, at $E/N\approx-1.3869$ and $-1.3861\,D$. At the lowest temperatures part of this bimodality comes from replicas that stay in one of the two states for the whole run, so the relative weights of the two peaks are not equilibrium weights.

\section{Correction of the SQUID data for the Py impurity}
\label{sm:impurity}

\begin{figure}[t]
\centering
\includegraphics[width=0.5\linewidth]{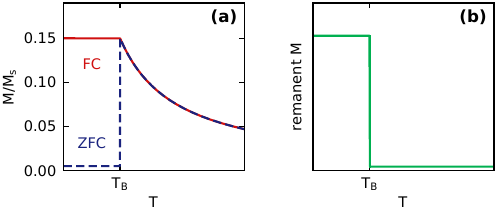}
\caption{\textbf{Sketch of the expected SQUID response of non-interacting, thermally active nanomagnets.} (a) FC (red) and ZFC (blue, dashed) magnetization: the ZFC curve stays near zero up to the blocking temperature $T_B$, joins the FC curve there, and both then decay as in a paramagnet. (b) Remanent magnetization after field cooling: constant up to $T_B$ and zero above. Interactions in the arrays of Fig.~\ref{figure:4} cause differences in the data.}
\label{figS:protocol}
\end{figure}

\begin{figure}[b]
\centering
\includegraphics[width=\textwidth]{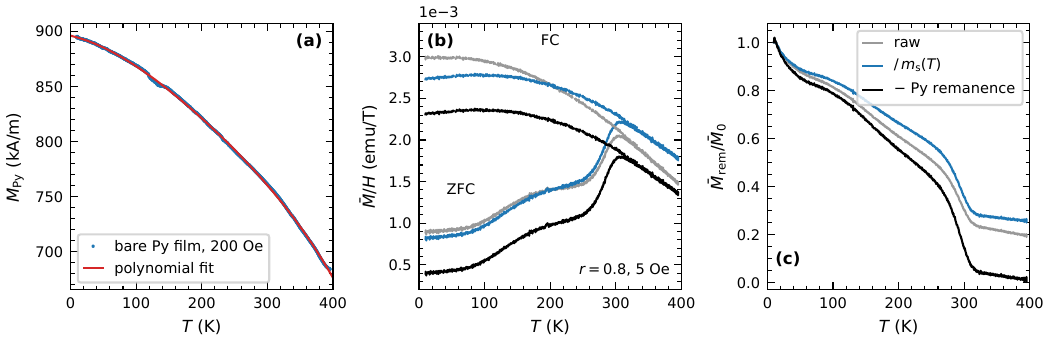}
\caption{\textbf{Corrections applied to the SQUID data}. (a)~Moment of the bare Py film in 200~Oe divided by its volume (dots) and polynomial fit used for $m_\mathrm{s}(T)$ (line). (b)~ZFC (solid) and FC (dashed) curves in 5~Oe: raw moment divided by 5~Oe (gray), after normalization by $m_\mathrm{s}(T)$ and division by $H_\mathrm{eff}$ (blue), and after subtraction of the Py offset $\chi_\mathrm{Py}$ (black, as in Fig.~\ref{figure:4}a). (c)~Remanent magnetization after cooling in 200~Oe, normalized to its 10~K value: raw (gray), divided by $m_\mathrm{s}(T)$ (blue) and after subtraction of the Py remanence $\kappa\chi_\mathrm{Py}$ (black, as in Fig.~\ref{figure:4}b). Correction parameters: $H_\mathrm{rem}=-3.0$~Oe, $\chi_\mathrm{Py}=4.2\times10^{-4}$~emu/T, $\kappa=1.3$~mT.}
\label{figS:correction}
\end{figure}

The ZFC-FC and remanent magnetization curves of Fig.~\ref{figure:4} were measured on the $r=0.4$, 0.6, 0.7, 0.8 series in an MPMS3 with the vibrating-sample option. In the ZFC-FC protocol, the sample is cooled in zero field to 10~K, a field of 5~Oe is applied and $M_\mathrm{ZFC}(T)$ is measured on warming to 395~K; $M_\mathrm{FC}(T)$ is then measured on cooling in the same field. For the remanent magnetization, the sample is cooled in 200~Oe from 395 to 10~K, the field is set to zero and $M_\mathrm{rem}(T)$ is measured on warming. Figure~\ref{figS:protocol} sketches the response of non-interacting, thermally active nanomagnets in these two protocols. We correct the raw data for three effects: the remanent field of the magnet, the temperature dependence of the saturation magnetization of Py, and a Py contribution that does not come from the patterned nanomagnets. Figure~\ref{figS:correction} shows the effect of each correction.

\paragraph*{Remanent field.}
After the magnet has been at high field, it adds a remanent field of a few Oe to the setpoint. The setpoint of the sequence carries an offset of $-2.5$~Oe, and the analysis divides the moment by the effective field
\begin{equation}
H_\mathrm{eff}=H_\mathrm{rep}-H_\mathrm{rem},
\label{eq:Heff}
\end{equation}
where $H_\mathrm{rep}$ is the field reported by the instrument and $H_\mathrm{rem}=-3.0$~Oe is the same for the four samples. This gives $H_\mathrm{eff}=5.1$ to 5.5~Oe for $r=0.6$--0.8, and 7.9~Oe for $r=0.4$, which was measured without the setpoint offset.

\paragraph*{Normalization.}
The nanomagnets in all samples have the same volume, but the saturation magnetization of Py decreases with temperature. We measured $M_\mathrm{Py}(T)$ of the bare Py film evaporated with the arrays (25~mm$^2$, 200~Oe, 10 to 395~K) and fitted a polynomial of degree five (Fig.~\ref{figS:correction}a). With a thickness of 2.4~nm this gives $M_\mathrm{Py}(0)=896$~kA/m and $M_\mathrm{Py}(300~\mathrm{K})=760$~kA/m. All moments are divided by
\begin{equation}
m_\mathrm{s}(T)=\frac{M_\mathrm{Py}(T)}{M_\mathrm{Py}(0)},\qquad \bar M(T)=\frac{M(T)}{m_\mathrm{s}(T)} ,
\label{eq:ms}
\end{equation}
with $m_\mathrm{s}(300~\mathrm{K})=0.85$. The remanent curves are in addition divided by their value at 10~K.

\begin{table}[t]
\caption{Correction parameters of the original sample series ($H_\mathrm{rem}=-3.0$~Oe, $\kappa=1.3$~mT). $\chi_\mathrm{Py}$ in $10^{-4}$~emu/T; $M_\mathrm{imp}=\kappa\chi_\mathrm{Py}$ is the remanent moment of the impurity in $10^{-7}$ emu; the last column gives it relative to the remanent moment of the array at 10~K.}
\label{tab:impurity}
\small
\begin{tabular}{cccc}
\hline\hline
$r$ & $\chi_\mathrm{Py}$ & $M_\mathrm{imp}$ & $M_\mathrm{imp}/M_\mathrm{rem}(10~\mathrm{K})$ \\
\hline
0.4 & 1.5 & 1.9 & 11\% \\
0.6 & 1.0 & 1.3 & 9\% \\
0.7 & 2.5 & 3.3 & 19\% \\
0.8 & 4.2 & 5.5 & 25\% \\
\hline\hline
\end{tabular}
\end{table}

\begin{figure}[b]
\centering
\includegraphics[width=\textwidth]{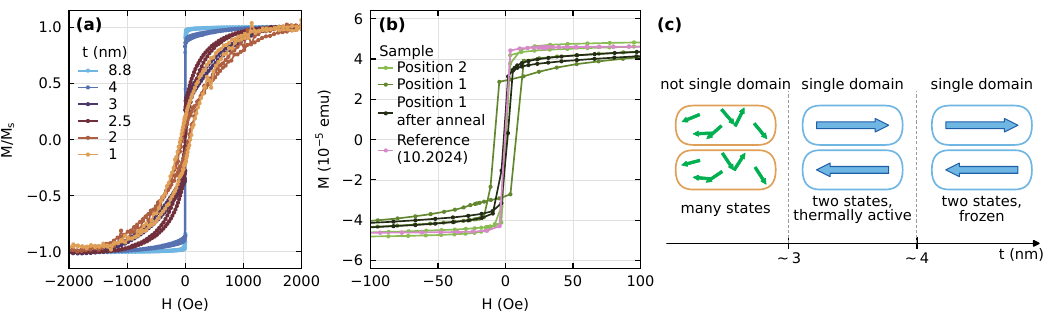}
\caption{\textbf{Py films and the magnetic states of a Py nanomagnet.} (a) In-plane $M(H)$ loops at 300~K of sputtered Py films of nominal thickness $t$ from 1 to 8.8~nm, each normalized to its saturation moment after subtraction of the substrate slope and of the remanent field of the magnet. The loops of the 1 and 2~nm films have a moment below $2\times10^{-6}$~emu and are amplified by the normalization. (b) $M(H)$ at 300~K of evaporated Py films used for the arrays (quartz-monitor reading 2~nm for the Reference and 4~nm for the 2025 film): film of the original sample series (Reference), then film evaporated at a separate time at two positions on the evaporator holder (Position 1 close to the arrays, Position 2 on the opposite side), and Position 1 after annealing. (c) Sketch of the magnetic states of a Py nanomagnet versus $t$. Below $t\approx3$~nm the loops of (a) are slanted, and an island has many internal states and is not single domain. Between about 3 and 4~nm an island is a single domain with two states, and for our $180\times60$~nm$^2$ islands these states flip thermally. Above about 4~nm the islands are Ising-like but frozen at room temperature. The thresholds are approximate.}
\label{figS:thickness}
\end{figure}
\paragraph*{Py impurity.}

After these two corrections, the remanent moment does not go to zero above the blocking temperature of about 310~K but stays on a plateau of 10\% to 26\% of its 10~K value, and the low-temperature ZFC curve is shifted upward. Continuous Py outside the arrays, or on the edges of the sample, explains both: after division by $m_\mathrm{s}(T)$ its contribution does not depend on temperature, it responds linearly to the small field and it keeps a remanent moment at zero field. We therefore subtract, for each sample, a constant $\chi_\mathrm{Py}$ from the susceptibility and a constant moment $\kappa\chi_\mathrm{Py}$ from the remanence:
\begin{align}
\frac{\bar M}{H}(T) &= \frac{M(T)}{m_\mathrm{s}(T)\,H_\mathrm{eff}}-\chi_\mathrm{Py},
\label{eq:chi}\\
\frac{\bar M_\mathrm{rem}(T)}{\bar M_0} &= \frac{M_\mathrm{rem}(T)/m_\mathrm{s}(T)-\kappa\chi_\mathrm{Py}}{M_\mathrm{rem}(10~\mathrm{K})-\kappa\chi_\mathrm{Py}} ,
\label{eq:rem}
\end{align}
with $\kappa=1.3$~mT for all samples and $\chi_\mathrm{Py}$ chosen so that the remanent plateau above the blocking temperature vanishes (Table~\ref{tab:impurity}). The impurity corresponds to a continuous film of 0.7\% to 3\% of the area of an array. The correction leaves the position of the ZFC maximum and the temperature at which the remanence collapses unchanged, since these do not depend on a constant offset.

\section{Influence of the bare Py magnetic properties and of the pattering quality on the thermally active behavior}
\label{sm:thickness}

The nanomagnets of the SQUID samples must fluctuate thermally around room temperature. This section shows that the right thermally active behavior is hard to obtain: it depends strongly on the Py thickness, on the quality of the evaporated film, and on the quality of the patterning.

We first characterized unpatterned Py films by SQUID magnetometry (MPMS3, in-plane field). We grew films of 1, 2, 2.5, 3, 4 and 8.8~nm by magnetron sputtering (base pressure $5\times10^{-8}$~mbar), since sputtering sets the thickness more precisely and more repeatably than our evaporator. Each film was capped with 2~nm of Al and cut to $5\times5$~mm$^2$. We measured $M(H)$ between $\pm2$~kOe at 300~K, after subtraction of the diamagnetic contribution of the substrate.

Figure~\ref{figS:thickness}a shows three groups. The 8.8 and 4~nm films have square loops with a single sharp switching step and a coercivity below 10~Oe. The 3 and 2.5~nm films still switch at about zero field, but this step carries only part of the signal; the rest approaches saturation only above about 700--1000~Oe, a sign of a magnetically hard or disordered fraction. The 2 and 1~nm films give a very weak signal and no single switching field. A nanomagnet behaves as a two-state Ising moment only above about 3--4~nm, and it is thermally active at room temperature only at the thin end of this range (Fig.~\ref{figS:thickness}c).

\begin{figure}[b]
\centering
\includegraphics[width=0.75\linewidth]{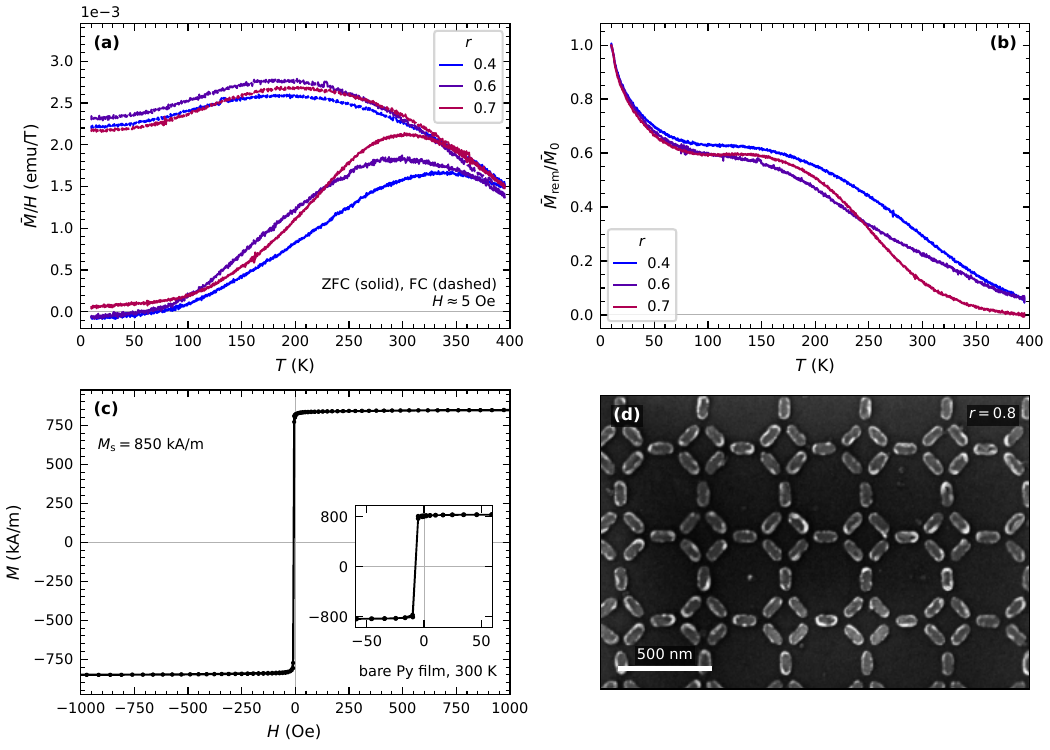}
\caption{\textbf{A good film is not enough: the patterning quality has a strong influence on the thermally active properties.} (a)~ZFC (solid) and FC (dashed) susceptibility in about 5~Oe for $r=0.4$, 0.6 and 0.7, corrected as in Fig.~\ref{figure:4}a (Sec.~\ref{sm:impurity}) with $H_\mathrm{rem}=-6.1$, $-4.7$, $-6.1$~Oe ($H_\mathrm{eff}=4.5$, 3.4, 4.5~Oe) and $\chi_\mathrm{Py}=0.7$, 0.8, $0.5\times10^{-4}$~emu/T. (b)~Remanent magnetization after cooling in 1000~Oe, divided by $m_\mathrm{s}(T)$, corrected for the Py remanence ($\kappa=1.3$~mT) and normalized to its 10~K value. (c)~Magnetization at 300~K of the bare Py film evaporated in the same run ($5\times5$~mm$^2\times3.6$~nm), after subtraction of the substrate slope; inset: zoom near zero field. The film saturates at $M_\mathrm{s}=850$~kA/m and switches in a single step, with no hysteresis resolved ($H_\mathrm{c}\lesssim3$~Oe); the step lies at $-7.5$~Oe because of the remanent field and sweep lag of the magnet. (d)~SEM image of the $r=0.8$ array of the same run (its SQUID data are not shown in a, b); islands about $124\times62$~nm$^2$, against a design of $125\times45$~nm$^2$.}
\label{figS:py2603}
\end{figure}

The ASIs arrays were evaporated, not sputtered, because lift-off works much better with directional evaporation. The properties of evaporated Py, however, change from run to run. Figure~\ref{figS:thickness}b compares the $M(H)$ loops at 300~K of the evaporated film of the arrays shown in the text (``Reference'', for the arrays of Fig.~\ref{figure:4}) with a film of the very similar thickness evaporated in November 2025 at two positions on the evaporator holder, one of which was also annealed for 30~min at 550~$^\circ$C. The saturation magnetization is similar in all films, but the low-field behavior, and with it the thermally active behavior, differs strongly. We attribute this to inconsistencies and small residual field in our evaporator setup.

Finally, a Py film with a nice square magnetization loop is not enough to get the desired thermally active behavior. We identified that the patterning quality also matters. Figure~\ref{figS:py2603} show measurements on a array made in a thicker Py film (to avoid the issues exposed above), but with a smaller lateral dimension so as to keep a similar amount of magnetic material per nanomagnet. We chose $c=450$~nm and lateral dimension of around 120 by 60 nms (Figure~\ref{figS:py2603}d), and evaporated around 3.6~nm of Py, with $r=0.4$, 0.6 and 0.7, corrected as in Sec.~\ref{sm:impurity}. The bare film of the same run is as soft as the film of the arrays of Fig.~\ref{figure:4}: it switches in a single step at 300~K, with no hysteresis resolved (Fig.~\ref{figS:py2603}c). The blocking temperature is similar in both series: the ZFC maximum lies between 300 and 340~K, against 311--316~K in Fig.~\ref{figure:4}. The blocking temperature is set by the energy barrier for reversing a nanomagnet, and this barrier comes from the magnetostatic (dipolar) energy of the nanomagnet, which scales as $M_\mathrm{s}^2$ times its volume and depends on its shape. A similar blocking temperature therefore means that the magnetic dipolar energy per nanomagnet must be similar in the two series. The single-nanomagnet energy scale cannot explain the differences between the two series. Yet every feature of the array data is broader. The ZFC maximum is a flat hump instead of a sharp peak. The remanent moment loses 40\% of its value by 100~K. We attribute this to local relaxation of the magnetization within each nanomagnet, made possible by multiple pinning sites on their edges. The difference within samples, driven by the change of interactions with the geometry, is also smaller in this series of samples. A spread of island shape and volume results in a spread of the blocking temperatures and the coupling strengths, and blurs the effect of the interactions in the magnetometry signal.

\section{Mapping charge order to a dimer model}
\label{sm:fragmentation}

\begin{figure}[tb]
\centering
\includegraphics[width=0.7\linewidth]{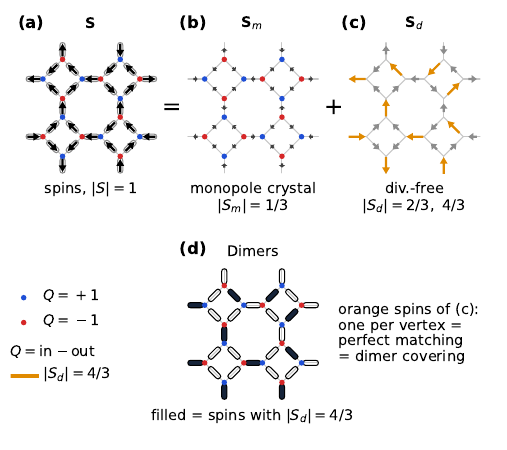}
\caption{\textbf{From charge order to dimers by moment fragmentation.} A charge-ordered Monte Carlo configuration ($r=0.6705$, $T=0.1884$, $2\times2$ unit cells; blue and red dots are vertices with $Q=+1$ and $-1$, where $Q$ is the number of incoming minus outgoing moments) (a) is the sum of (b) a monopole crystal of moments of length $1/3$ pointing from the $-1$ to the $+1$ vertices and (c) a divergence-free fragment, with lengths $2/3$ (grey) and $4/3$ (orange). (d) The moments of length $4/3$ (filled) are one per vertex and form a dimer covering. Arrow lengths are proportional to the fragment lengths.}
\label{figS:frag}
\end{figure}

Every vertex of the CaVO lattice joins three nanomagnets, so its magnetic charge $Q_v=(\#\text{in})-(\#\text{out})$ is $\pm1$ whenever the ice rule holds. In the charge-ordered phases, $Q_v=+1$ (two in, one out) and $Q_v=-1$ (one in, two out) alternate on the two sublattices of the bipartite vertex graph, so every nanomagnet joins a $+1$ and a $-1$ vertex. Following the fragmentation construction~\cite{brooks-bartlettMagneticMomentFragmentationMonopole2014,canalsFragmentationMagnetismArtificial2016}, we apply a discrete Helmholtz decomposition to the magnetization field. Each moment $\mathbf{S}_i$ (of unit length) splits into a divergence-full and a divergence-free fragment:
\begin{equation}
\mathbf{S}_i=\mathbf{S}^{m}_i+\mathbf{S}^{d}_i .
\label{eq:frag}
\end{equation}
The monopole fragment $\mathbf{S}^{m}_i$ has length $1/3$ and points from the $-1$ vertex to the $+1$ vertex of the nanomagnet. The three fragments at a vertex then all point in, or all point out, so they carry the full charge $Q_v$ and form an all-in/all-out crystal (Fig.~\ref{figS:frag}b). The remainder $\mathbf{S}^{d}_i=\mathbf{S}_i-\mathbf{S}^{m}_i$ has no divergence. At a $Q_v=+1$ vertex, the two moments that point in are parallel to $\mathbf{S}^{m}$ and keep a length $1-1/3=2/3$. The moment that points out is opposite to $\mathbf{S}^{m}$ and gets a length $1+1/3=4/3$. The charge of $\mathbf{S}^{d}$ is $2\times\frac23-\frac43=0$, as required, and the same holds at a $Q_v=-1$ vertex with all directions reversed (Fig.~\ref{figS:frag}c).

Each vertex thus carries exactly one nanomagnet with $|\mathbf{S}^{d}|=4/3$. Because of this strong constraint, the $\mathbf{S}^{d}$ field can be mapped to a dimer model on the links of the CaVO lattice (Fig.~\ref{figS:frag}d), where by definition each vertex must have exactly one dimer around itself. The ordered monopole crystal $\mathbf{S}^{m}$ is the same in every charge-ordered configuration, and all the remaining disorder sits in the dimer covering. The dimer solid, the dimer paramagnet and the Coulomb phase of the main text therefore differ only in which coverings the system samples. Figure~\ref{figS:frag} shows the construction on a Monte Carlo configuration at $r=0.6705$ and $T=0.1884$ (Coulomb phase), cut to $2\times2$ unit cells.

\vspace{1cm}

\end{document}